# SPECTRAL SIMILARITY OF UNBOUND ASTEROID PAIRS

Stephen D. Wolters [a,1], Paul R. Weissman [a], Apostolis Christou [b], Samuel R. Duddy [c], Stephen C. Lowry [c]

[a] *Planetary Science Section, Jet Propulsion Laboratory, California Institute of Technology, 4800 Oak Grove Drive, Pasadena CA 91109, USA;*
[b] *Armagh Observatory, College Hill, Armagh, BT61 9DG, UK;*
[c] *Centre for Astrophysics and Planetary Science, School of Physical Science, University of Kent, Canterbury, CT2 7NH, UK.*

e-mail: stephen.wolters@open.ac.uk
e-mail addresses of co-authors: paul.r.weissman@jpl.nasa.gov, aac@arm.ac.uk; s.duddy@kent.ac.uk; s.c.lowry@kent.ac.uk

---

[1] Current Affiliation: The Open University, Walton Hall, Milton Keynes MK7 6AA, stephen.wolters@open.ac.uk

**Abstract**

Infrared spectroscopy between 0.8 and 2.5 microns has been obtained for both components of three unbound asteroid pairs, using the NASA-IRTF with the SpeX instrument. Pair primary (2110) Moore-Sitterly is classified as an S-type following the Bus-DeMeo taxonomy; the classification for secondary (44612) 1999 RP27 is ambiguous: S/Sq/Q/K/L-type. Primary (10484) Hecht and secondary (44645) 1999 RC118 are classified as V-types. IR spectra for Moore-Sitterly and Hecht are each linked with available visual photometry. The classifications for primary (88604) 2001 QH293 and (60546) 2000 EE85 are ambiguous: S/Sq/Q/K/L-type. Subtle spectral differences between them suggest the primary may have more weathered material on its surface. Dynamical integrations have constrained the ages of formation: 2110-44612 > 782 kyr; 10484-44645 = 348 (+823,-225) kyr; 88604-60546 = 925 (+842,-754) kyr. The spectral similarity of seven complete pairs is ranked in comparison with nearby background asteroids. Two pairs, 17198-229056 and 19289-278067, have significantly different spectra between the components, compared to the similarity of spectra in the background population. The other pairs are closer than typical, supporting an interpretation of each pair's formation from a common parent body.

# 1 Introduction

Vokrouhlický and Nesvorný (2008) discovered pairs of asteroids with osculating orbital elements too similar to have likely arisen coincidentally from random fluctuations of local background asteroid population density. Pravec and Vokrouhlický (2009) developed a method, applicable to many pairs, for estimating the probability of them arising coincidentally between genetically unrelated bodies. This was formalized as the ratio $P_2/N_p$, where $P_2$ is the number of pairs expected in a defined broad orbital zone (e.g. inner main belt) and $N_p$ is the number of pairs found within a distance in osculating orbital element space, normalized by a factor $R_0$ that encodes the local population density (which can be non-constant). Hence the most statistically significant pairs were found to be the closest (small distance) or those that lie in low-density regions of space (large $R_0$). If the long term dynamical behaviour is also found to be similar, i.e. if the components have nearly identical proper orbital elements (proper element distance < 10 m/s) then Pravec and Vokrouhlický considered such a pair to be reliable; they identified over 80 of these.

If $P_2/N_p$ < 0.01 and Pravec and Vokrouhlický's method is valid, one can be >99% confident that a pair's components are genetically related, i.e. formed from a common parent body. Backward dynamical integrations can provide an independent method of deducing that a pair is genetically related. Virtual clones of the pairs are produced from the covariant matrix of orbital position uncertainty and a range of assumed Yarkovsky drift values. Convergence of two clones can be defined as reaching a relative velocity less than the escape velocity ($dv < v_{esc}$) and a minimum distance less than the Hill Sphere radius ($dr_{min} < R_{Hill}$). Given a low $P_2/N_p$, we do not need to be concerned if only a small fraction of the simulated clones converge. The age distribution of converged clones can then be used to estimate the pair formation age. Pravec et al. (2010) performed such a simulation for 35 pairs by performing backward integrations over 500 kyr or 1000 kyr, estimating ages where convergence was reached and constraining the minimum ages if it was not. It is important to note that earlier than the Lyapunov timescale, the relative orbital longitude between clones is randomized, hence the minimum distance measure for a given clone pair no longer yields meaningful positional information. However, the relative velocity and orbital convergence metric remain reliable. For example, the pairs studied in Duddy et al. (2013) all formed earlier than the longitude randomization timescale, hence the convergence constraint was relaxed to require minimum velocity only.

The Pravec and Vokrouhlický (2009) method of identifying reliable pairs is not all-inclusive and leaves out several likely pairs; for example, 6070-54827 has an extremely precise age from backward dynamical integrations of 17.22 ± 0.22 kyr (Vokouhlický and Nesvorný, 2009). Unlike that pair, most cannot be definitively established from dynamical integrations due to longitude randomization. Hence, it would be helpful if another line of evidence can be established to test whether a given pair has a common origin. Duddy et al. (2012) found that 7343-154634 had very similar visual spectra (0.40-0.92 μm). Duddy et al. (2013) found two further pairs, 1979-13732 and 19289-278067, with similar visual spectra (0.50-0.92 μm). However, pair 17198-229056 appeared dissimilar, with a deeper one micron absorption band for the secondary. Moskovitz (2012) obtained *BVRI* photometry for eleven pairs and found that they were also similar; in 98% of cases, when the observed pairs were compared to random samplings of two nearest neighbours taken from the Sloan Digital Sky Survey (SDSS), the pairs were closer in colour between their components.

The gentle separation of the components implied by the similarity of proper orbital elements and convergences $dv < v_{esc}$ suggests that rotational fission (Walsh et al. 2008) is the most plausible agent of their formation, with the YORP effect (Rubincam, 2000) the likely mechanism behind the spin-up. After the components first separate, the secondary will be an initially unstable satellite in orbit about the primary. Scheeres (2009) found that if the ratio of the secondary mass with the primary $\Delta Q$ is less than about 0.2, then after brief dynamical interactions the secondary can extract enough rotational energy from the primary to escape the gravitational potential of the system and enter an independent orbit about the Sun. Pravec et al. (2010) performed a survey of 35 pairs and reported longer rotation periods of primaries as the 0.2 mass ratio was approached, providing strong observational evidence for formation of asteroid pairs through rotational fission.

We report the first results of a survey of asteroid pairs started in February 2012 with the NASA Infrared Telescope Facility (IRTF) on Mauna Kea, Hawaii. In this work we present spectra of the three complete pairs observed so far and undertake new dynamical studies. These pairs are identified in Pravec and Vokrouhlický (2009) using the criteria described above. We aim to assess the evidence of pair formation from a common parent body by measuring the spectral similarity between the pair components, together with the previous complete pairs observed by Duddy et al. (2012, 2013). A forthcoming paper will report spectra for asteroids where we have only observed the primary pair component. In that work, we will consider the evidence for resurfacing from pair formation and subsequent rapid space weathering, or lack thereof.

## 2 Observations and Data Reduction

We used the SpeX instrument (Rayner et al., 2003) in prism mode with a 0.8'' slit to obtain 0.82-2.49 µm spectra of pairs 2110-44612, 10484-44645, and 88604-60546. Images were taken in AB nod pairs, with each nod having an exposure time of 120 s. Circumstances of the observations are given in Table 1. The reduction procedure was performed using a semi-automated pipeline written in Python with the Pyraf module implemented to interface with the appropriate IRAF routines. Images had bad pixel masks applied. They were flat-fielded, A-B image pairs subtracted, the wavelength dispersion corrected and linearised to 0.05 µm bins, trimmed, shifted and scaled. A combined image was created from all useable beam images, then the spectral aperture traced and spectrum extracted.

For the asteroid images, the spectrum was also extracted for each individual frame. Each individual spectrum was ratioed with the solar analogue star to produce relative reflectance values and manually inspected for contamination. In addition the pipeline automatically ranked spectrum quality based on signal level, and a noise level calculated from residuals to polynomial fits over three sections within the spectrum. The noisiest frames were automatically removed one by one from the sequence, and a new combined spectrum created.

The combined asteroid and solar analogue star spectra are telluric-corrected by an IDL routine (S. J. Bus, private communication) that best-fits an ATRAN model atmospheric spectrum to the spectra over two overlapping wavelength ranges, through minimisation of residuals to polynomial fits (4th order for 1.12-1.75 µm, 5th order for 1.52- 2.49 µm).

This allows an empirical correction of differences in precipitable water vapour level in the atmosphere between the solar analogue and asteroid observations. A by-product of this routine is that bad relative reflectance values where telluric correction was unsuccessful are identified and labelled. These are removed from the given Figures and skipped for smooth-spline creation during spectral-type classification. Another advantage of this method is that solar analogue stars do not need to be observed at very close angular distances in the sky, although where possible we observed several analogues covering the range of airmass traversed by the asteroid.

**Table 1: Observational Circumstances**

| Asteroid | p/s | date (dd/mm/yyyy) | start end | tot. im. | $\chi$ range | V (mag) | $\alpha$ (°) | PW (mm) | Solar analogue | $\chi$ |
|---|---|---|---|---|---|---|---|---|---|---|
| (2110) Moore-Sitterly | p | 02/03/2013 | 08:49 09:44 | 24 (24) | 1.062 1.167 | 17.0 | 1.0 | 0.67 | FS16 *SA102-1081* SA105-56 | 1.01 *1.10* 1.16 |
| (44612) 1999 RP27 | s | 31/08/2012 | 10:47 11:20 | 14 (9) | 1.107 1.111 | 16.8 | 5.2 | 2.4 | SA115-271 *SA93-101* SA93-101 | 1.10 *1.09* 1.09 |
| (10484) Hecht | p | 01/04/2013 | 07:21 10:00 | 44 (5) | 1.152 1.441 | 16.9 | 5.0 | 1.4 | *FS16* | *1.01* |
| | | 14/05/2013 | 08:14 09:15 | 26 (21) | 1.217 1.442 | 17.9 | 20.4 | 2.1 | SA102-1081 *SA105-56* | 1.23 *1.19* |
| (44645) 1999 RC118 | s | 19/03/2012 | 07:54 08:57 | 22 (22) | 1.321 1.612 | 17.6 | 5.1 | 0.49 | HD86627 *HD91859* | 1.54 *1.39* |
| (88604) 2001 QH293 | p | 05/02/2012 | 13:00 13:56 | 24 (24) | 1.214 1.455 | 16.8 | 6.1 | 0.42 | HD88371 *HD88371* HD88371 | 1.03 *1.27* 1.49 |
| (60546) 2000 EE85 | s | 15/02/2013 | 12:45 14:05 | 26 (25) | 1.341 1.928 | 17.7 | 5.2 | 1.1 | SA105-56 SA102-1081 *SA102-1081* | 1.07 1.43 *1.71* |

Notes: p/s = primary or secondary; start to end time not corrected for light travel time; tot. im. = Number of images taken (in parentheses: number in combined spectrum after manual and/or automatic frame removal); $\chi$ = airmass; *V* = apparent visual magnitude, taken from JPL Horizons web service; $\alpha$ = solar phase angle; PW = precipitable water, fitted value over 1.25-2.25 micron reported by telluric correction routine. Solar analogue standard star in italics used for final relative reflectance spectrum, followed by the airmass of its observation.

The noise in the final spectrum is measured from residuals to polynomial fits, and then the process is repeated for the next combined image with the noisiest individual exposure removed. This is repeated until we are confident that the optimum combined image that can result from this process has been captured. A noisy image might result from poor centering in the slit or from a spike in water vapour. In some cases, the optimum S/N is attained without removing any frames, usually in good conditions for brighter targets. The final relative reflectance spectrum obtained by ratio with each solar analogue was compared. In each case, differences were negligible.

Asteroid (10484) Hecht was observed twice, on 1 April 2013 and 14 May 2013. Conditions were poor on 1st April, with some cloud and > 3'' seeing, and many frames were unusable. However, the spectra of 1st April are consistent below 2 microns with those obtained under better conditions (Figure 1). Asteroid (60546) 2000 EE85 was at a high airmass at the end of its observation sequence. To check its spectral shape was not affected, we split the images into lower and higher airmass sequences; the spectra are consistent (Figure 2).

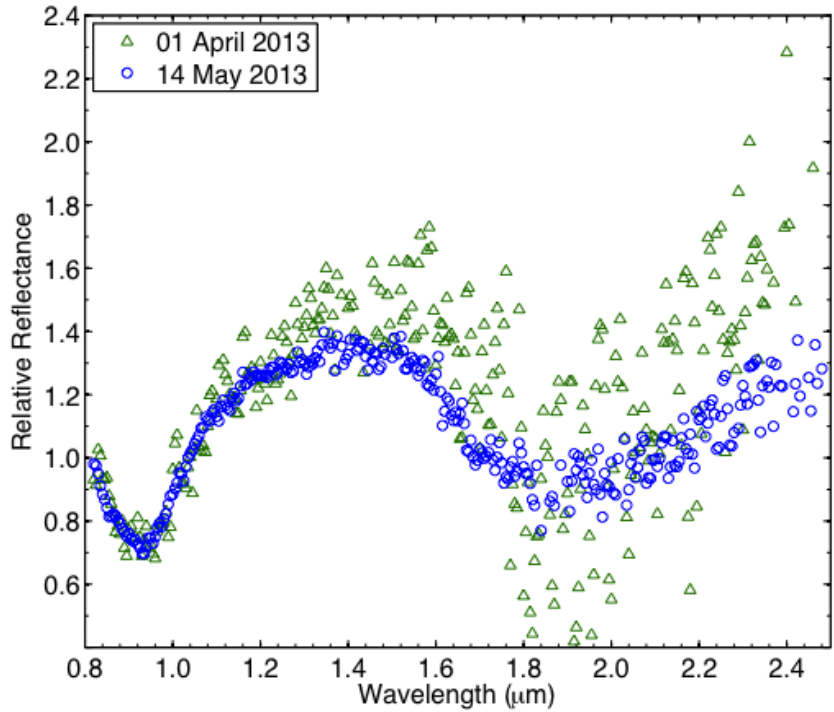

Fig. 1. SpeX relative reflectance spectra of (10484) Hecht.

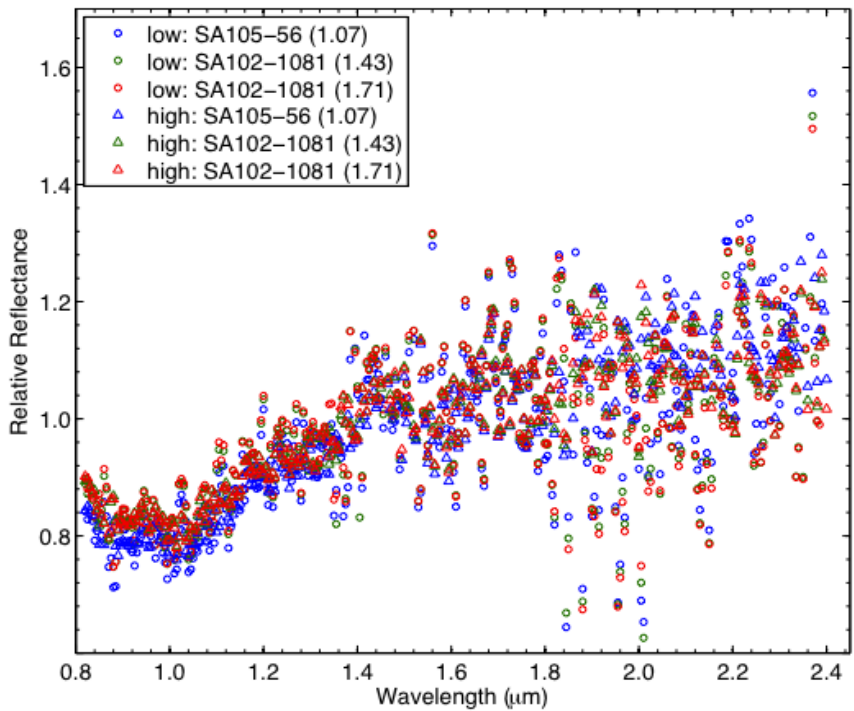

Fig. 2 SpeX relative reflectance spectra of (60546) 2000 EE85 taken on 15 February 2012, split into two 13 image groups: a lower airmass sequence (low : airmass = 1.352-1.650) and a high airmass sequence

(high: airmass = 1.671-1.983). Relative reflectance spectra for each solar analogue is also presented. Data > 2.4 microns had a poor telluric correction and is removed.

## 3 Results

The relative reflectance spectra are given in Figure 3. For asteroids (2110) Moore-Sitterly and (10484) Hecht, the spectra are linked with available *ugriz* photometry taken from the 4th SDSS Moving Object Catalog and compiled in Willman et al. (2010). Moore-Sitterly is also linked with *BVRI* photometry taken from Ye (2011) and Moskovitz (2012). Following Moskovitz (2012), SDSS magnitudes are converted to relative reflectance values from the bands centered at 0.35, 0.47, 0.62, 0.75 and 0.89 microns, using solar colours: $u\text{-}g$ = 1.43 ± 0.05, $g\text{-}r$ = 0.44 ± 0.02, $r\text{-}i$ = 0.11 ± 0.11, $i\text{-}z$ = 0.03 ± 0.02. We normalized at 0.55 microns from linear interpolation of the *g*- and *r*-band reflectances. *BVRI* magnitudes have band centres of 0.45, 0.55, 0.64 and 0.85 microns and solar colours used are: $B\text{-}V$ = 0.65, $V\text{-}R$ = 0.36, and $R\text{-}I$ = 0.32.

Spectra were classified following the Bus-DeMeo taxonomy (DeMeo et al., 2009), applying the VIS+IR or IR-only scheme, as appropriate. Classifications are given in Table 3, along with new classifications for the spectra for the complete pairs obtained at visual wavelengths in Duddy et al. (2012, 2013). Since the Bus-DeMeo taxonomy is an extension into the infrared of the feature-based taxonomy introduced by Bus (1999), and as such is designed to be consistent with it, the visual spectra have been reclassified following the Bus taxonomy (Bus and Binzel, 2002b), applying the decision tree given in Bus (1999).

*Figure 3*
(a)

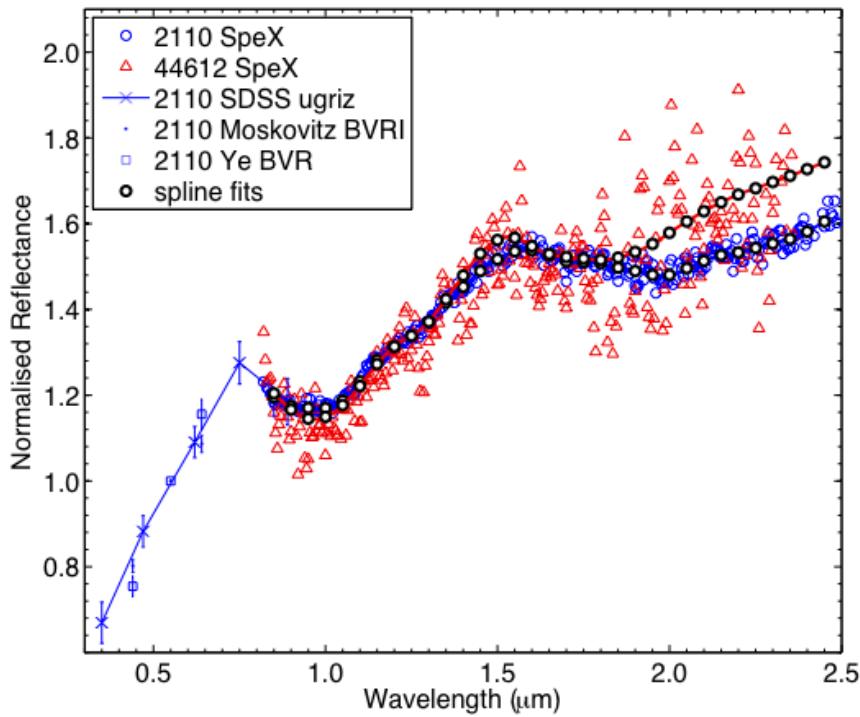

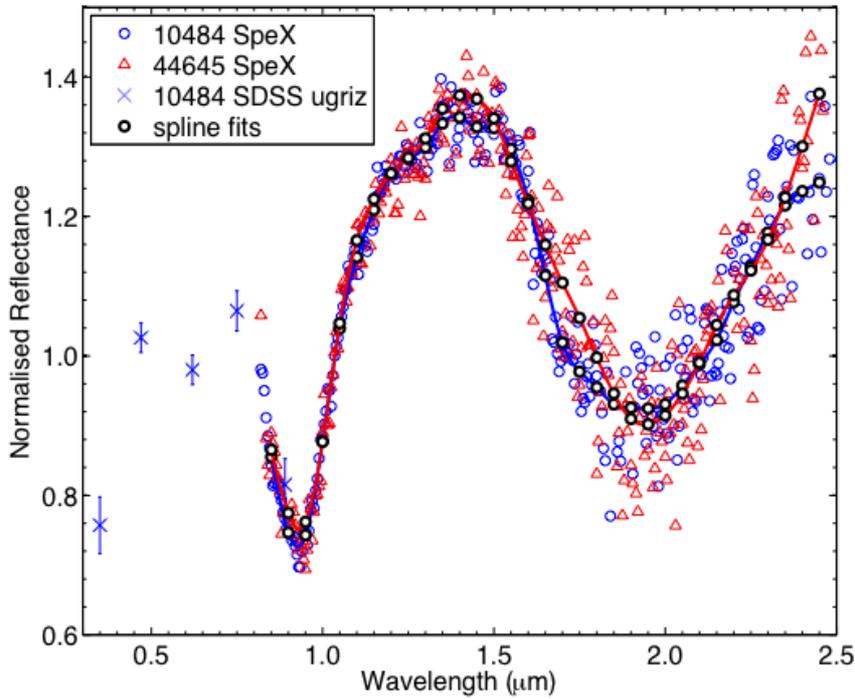

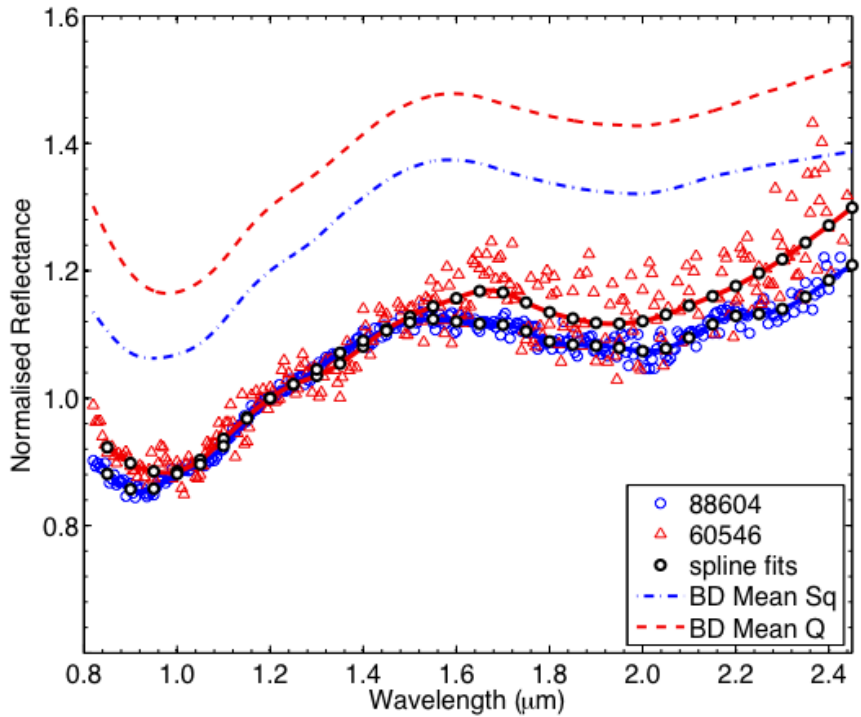

Fig. 3. 0.82-2.49 micron relative reflectance spectra obtained with SpeX, Blue circles are the primary asteroid, red triangles are the secondary. Black circles within solid lines are smoothed cubic splines (Reinsch, 1967) sampled at 0.05 micron intervals from 0.85-2.45 microns. (a) 2110-44612: smoothing parameter set to 1.0, *BVRI* and *ugriz* photometry convolved with (2110) Moore-Sitterly IR spectrum, normalised at 0.55 microns; (b) 10484-44645: 14 May 2013 data used for (10484) Hecht, smoothing parameter set to 0.1, normalised at 1.2 microns; (c) 88604-60546: smoothing parameter set to 1.0, normalised at 1.2 microns; also plotted are mean Sq and Q spectra from DeMeo et al. (2009), offset by +0.2 and +0.3 respectively, for clarity.

We found the five nearest neighbours to each pair for which there is exists a recorded infrared spectrum, taken over the range 0.85-2.45 micron, and in the same taxonomic complex, available in the Small Main-Belt Asteroid Spectroscopic Survey catalogue (SMASS, smass.mit.edu). We define the nearest asteroid as that listed in the catalogue with the smallest distance in proper orbital element space defined by the metric given in Zappalà et al. (1990), using the synthetic proper elements given by the AstDys web service (Figure 4). The neighbours are compared with our pair spectra in Figure 5. The neighbour spectra are smoothed with a cubic spline (Reinsch, 1967) and re-sampled at 0.05 micron intervals. The square of the difference between the reflectance values at each point is summed to provide a measure of the spectral similarity. We performed the same test on the pairs reported in Duddy et al. (2012, 2013), with nearest SMASS visual spectra (Xu, 1994; Xu et al., 1995; Bus, 1999; Bus and Binzel, 2002a) selected as neighbours. For each group of seven objects, i.e. the pair plus the five nearest neighbours, there are 21 unique pairings. The resulting ranking (Table 3) measures the spectral similarity of the pair compared to other asteroids, of the same taxonomic complex, and broadly in the same orbital phase space. However, we note that the closest background asteroids in the SMASS catalogue for 2110-44612 and 88604-60546 are fairly distant in orbital phase space in comparison with the other pairs. Therefore the ranking for these pairs should be treated with caution.

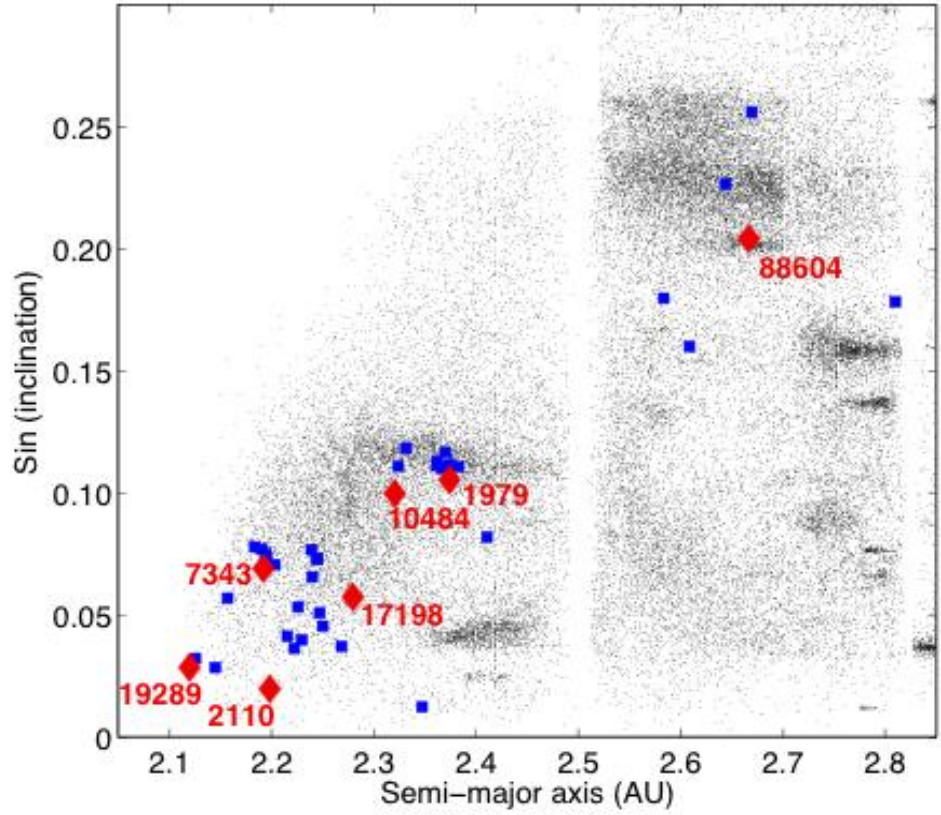

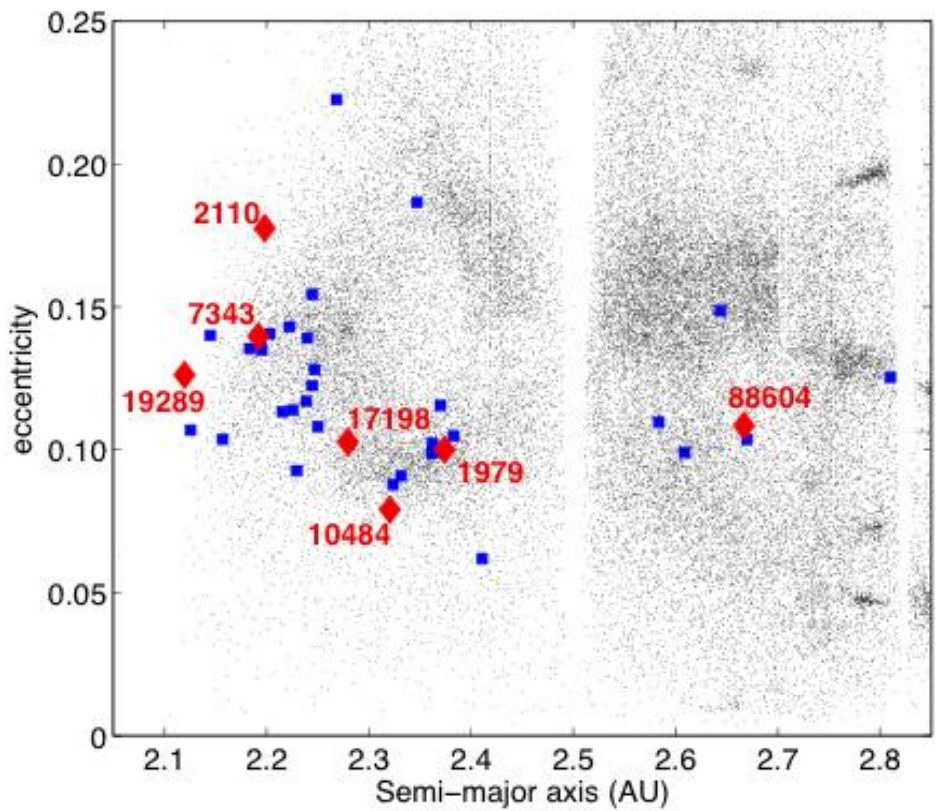

Fig. 4. Synthetic proper orbital elements taken from AstDys web service. Red diamonds: asteroid pairs observed in this work and Duddy et al. (2012, 2013) (primaries labelled); blue squares: background asteroids used for spectral similarity comparison tests; black dots: numbered asteroids with absolute visual magnitude $H$ < 15.

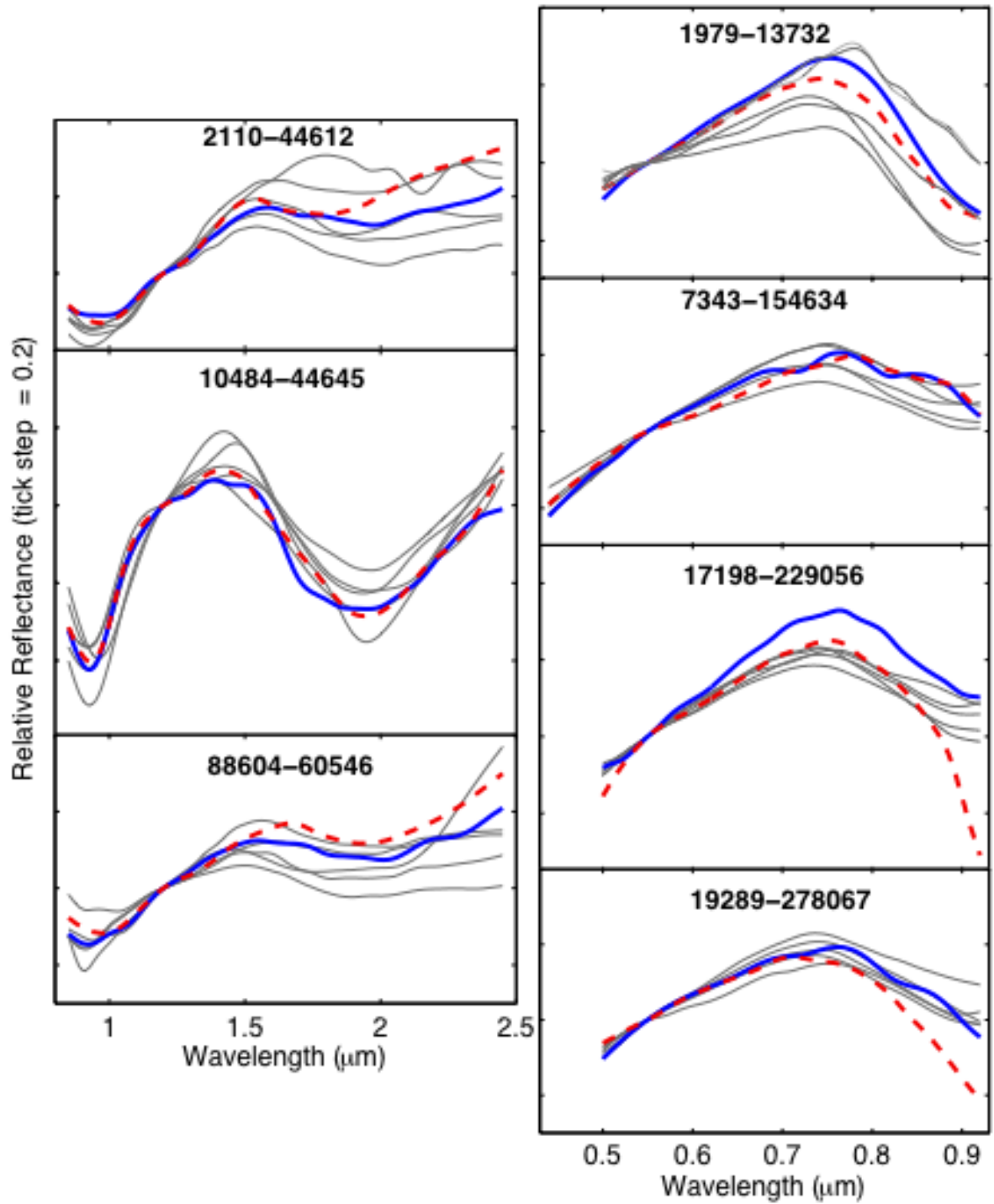

Fig. 5: Spectra of primary (thick blue line) and secondary (thick dashed line) pair components with nearest neighbours available in SMASS catalogue (thinner grey lines). Visual spectra (right) are smoothed with a cubic spline and re-sampled at 0.01 micron intervals, resulting in a 43 point spectrum (49 for 7343-154634); infrared spectra (left) are re-sampled at 0.05 micron intervals, resulting in a 33 point spectrum.

## 4. Dynamics

The numerical simulations of the asteroids' dynamical evolution were set up in the same way as with Duddy et al. (2013). We carried out trial simulations with the major planets as perturbers, with and without the large asteroids Ceres and Vesta as massive bodies perturbing the asteroid clones. We found no significant differences between the two cases and baselined the latter model for the main integrations. Two sets of integrations were carried out. The first set was run for 2000 kyr with a 1000 year output time step and the second set for 200 kyr with a 100 year step. Both sets were tested for pairs that achieve $dv < v_{esc}$ and, at the same time, $dr_{min} < R_{hill}$. For these purposes, the evolution of $dr_{min}$ is tracked every 10 years in the first set and every 1 year in the second set of integrations by approximating the relative motion between the components with the Hill equations of motion (Duddy et al., 2013). The second set of integrations serves to identify solutions that may have been missed by the first, due to the factor of ten coarser output time step. This is feasible near the beginning of the 2000 kyr integration time span where the information regarding positional convergence in the data has not yet been destroyed due to the chaotic nature of the asteroid orbits.

As in Duddy et al. (2013), we have used the evolution of the relative mean longitude $dl$ to indicate how far from the start of the integration the relative position between the pair components can remain tractable. In all three cases studied here, this timescale was in the range 100-150 kyr. While it is still meaningful to explore the possible minimum distance for those times beyond dl randomization, their time distribution may widen to the point that individual clusters of solutions become difficult to identify, and generally occur for a low fraction of clones, e.g. 10 out of 10 000 combinations.

Effective spherical diameters $D_{eff}$ and geometric visual albedos $p_V$ were obtained from the preliminary results of the NEOWISE survey (Masiero et al., 2011), and are given in Table 3. For (60546) 2000 EE85, we estimate the diameter assuming it has the same albedo as (88604) 2001 QH293, and that $H_V \sim 14.4$ (Pravec and Vokrouhlický, 2009). Albedos are converted to visual bolometric Bond albedo assuming phase parameter $G = 0.24$ for S-complex pairs 2110-44612 and 88604-60546, and $G = 0.43$ for V-type pair 10484-44645 (Warner et al., 2009). $v_{esc}$ and $R_{Hill}$ for convergence assessment were calculated by summing together the volumes of the two components and assuming a density of 2500 kg m$^{-3}$, while 1000 kg m$^{-3}$ was assumed to calculate the boundary Yarkovksy drift rate when generating clones, in order to capture all plausible values (both as surface density to determine the thermal response part of the model, and bulk density to determine the acceleration).

For all three cases and for both sets of integrations, none of the clone pairs satisfied both the orbital and positional convergence criteria simultaneously (with a few exceptions, see Section on 10484-44645). This is either because such convergences, particularly positional, lie too far in the past to be recognisable, or because the asteroids are not genetically related after all. Consequently, we resorted to using only the orbital velocity convergence criterion to constrain the formation age. The evolution of minimum distances $dr_{min}$ and relative orbital speeds $dv$ are shown in Fig. 6, and histograms of the number of pairs that achieved orbital velocity convergence in Fig. 7 and 8. To consider our results in context, we derived median ages and 90%/95% confidence limits of the $dv$ distribution (Table 2). We updated the dynamical analysis for the pairs in Duddy et al. (2012, 2013) to produce similar histograms and measure the median age and confidence limits in a consistent manner. These are also given in Fig. 8 and Table 2.

*2110-44612*

The component orbits remain too different to allow any of the clone pairs to converge for up to 500 kyr in the past (*dl* randomizes after 120 kyr). Beyond that point, the frequency of solutions follows a distribution punctuated by a sharp maximum in the interval 800-900 kyr. We caution that since a significant fraction of the distribution appear to lie beyond the 2 Myr sampled by our integrations and the counts show no sign of petering off at that point, these statistical descriptors of said distribution are lower limits themselves. Consequently, this pair may well be older than 2 Myr.

*10484-44645*

The relative velocity for a good fraction of pairs falls below $v_{esc}$ at 30-60 kyr and again after ~100 kyr. At the same time, the minimum distance becomes comparable to $R_{Hill}$. However, no pairs were found where the two conditions are satisfied simultaneously. Since the relative longitude dl randomizes after 100 kyr, this feature occurs early enough in the history of the actual pair to expect a valid solution to be present in the data. Therefore we regard it as spurious and unrelated to the splitting of the progenitor.

The relative velocity statistics alone shows a rapidly decreasing trend, which suggest an age much younger than 2000 kyr. The *dv* histogram corresponding to the 200 kyr integrations allows us to look at the convergence statistics at high resolution; the peak at 30-60 kyr is evidently related to the minimum identified in Fig 6; here we have repeated our search for pairs that satisfied both conditions. A few clone pairs do achieve positional convergence at the same time as orbital convergence in this interval. However, due to their small number, ~10 out of 100 × 10201 possible pairs tested within that time bin, and the relatively early occurrence in the integration which should result in a much higher success rate for the clones (c.f. Pravec et al, 2010), we are confident this is a spurious feature, rather than actual evidence for an age as short as 40 kyr. Considering only the counts from 70 kyr onwards then yields the age statistics in Table 2.

*88604-60546*

The relative velocity for most clone pairings show minima at the escape velocity threshold at 10-15 kyr and again at 70-100 kyr. These coincide with respective minima in $dr_{min}$ comparable to the Hill sphere radius. However, as in the case of 10484-44645, a search for full convergence in both sets of simulations yielded negative results (*dl* randomizes after 150 kyr). Therefore, the *dv* statistics for > 20 kyr are used to calculate the ages in Table 2. Inspection of the velocity histogram for the 2000 kyr simulations shows two maxima within the first 400 kyr and a smooth distribution of pairs with *dv* < $v_{esc}$ after that. Most of the power in the first peak comes from the 10-20 kyr minimum in *dv* while lesser peaks visible in the statistics of the 200 kyr simulations also do not show any simultaneous orbital and positional convergence. A second, broader peak is visible at 200-500 kyr. Finally we note that, although the distribution appears truncated at 2000 kyr, in contrast to the case of 2110-44612, this occurs well after the mode and is not likely to significantly bias the statistics towards younger ages.

**Table 2: Age estimates (in kyr) for different confidence limits**

| Pair | L95 | L90 | M | U90 | U95 |
|---|---|---|---|---|---|
| 2110-44612 | 643 | 782 | 1383 | 1935 | 1964 |
| 7343-154634 | 410 | 442 | 982 | 1862 | 1919 |
| 10484-44645 | 101 | 123 | 348 | 1173 | 1436 |
| 11842-228747 | 92 | 94 | 301 | 1205 | 1436 |
| 17198-229056 | 95 | 113 | 285 | 1060 | 1360 |
| 19289-278067 | 525 | 565 | 1189 | 1718 | 1834 |
| 88604-60546 | 92 | 171 | 925 | 1767 | 1858 |

Key: [L95, U95] = lower and upper 95% confidence limits (i.e. 2.5% of the distribution outside the limit at each tail); [L90, U90] = lower and upper 90% confidence limit (i.e. 5% of the distribution outside the limit at each tail); M = median age.

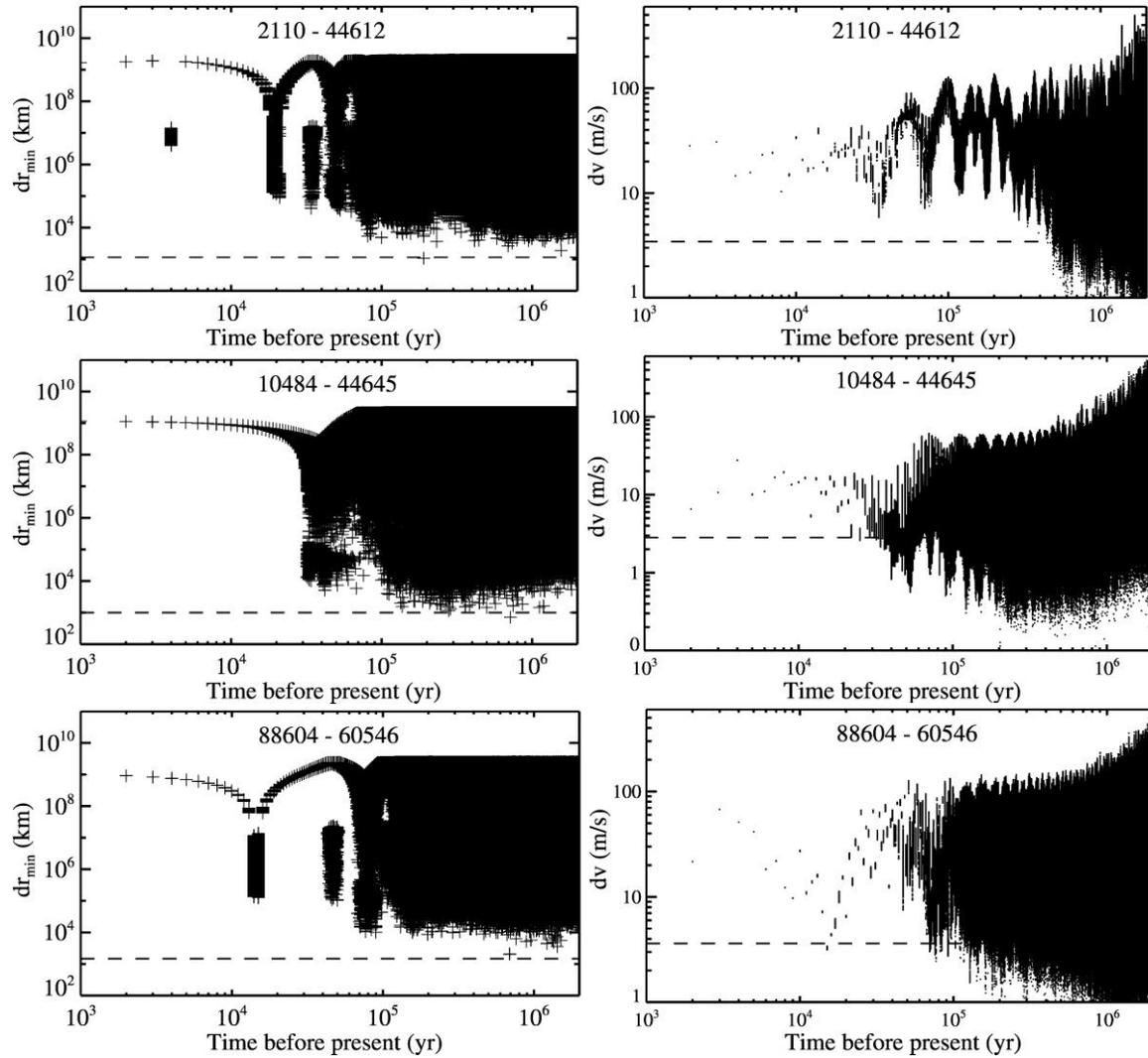

Fig 6. Left panel: Time evolution of minimum orbital distance $dr_{min}$ between possible primary and secondary component clone pairings, for the 2000 kyr integrations. To minimise clutter, only every other clone has been considered (2601 pairings). The dashed line denotes the Hill radius $R_{Hill}$ of the pair's progenitor.

Right panel: Time evolution of relative orbital speed $dv$ between all 10201 possible clone pairings are considered. The dashed line denotes the escape velocity $v_{esc}$ at the progenitor's surface.

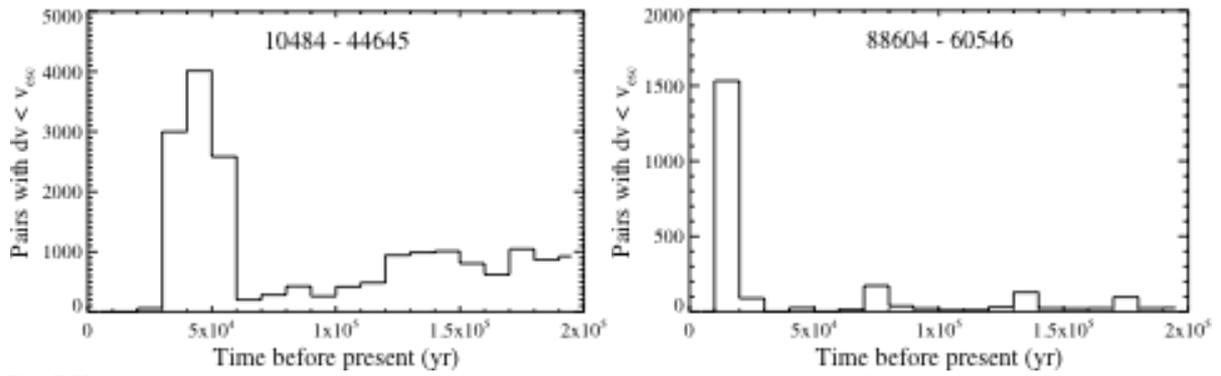

Fig. 7 Histogram of the number of clone pairings for which d$v$ < $v_{esc}$ during 200 kyr integration (there are no pairings in this time span for 2110-44612).

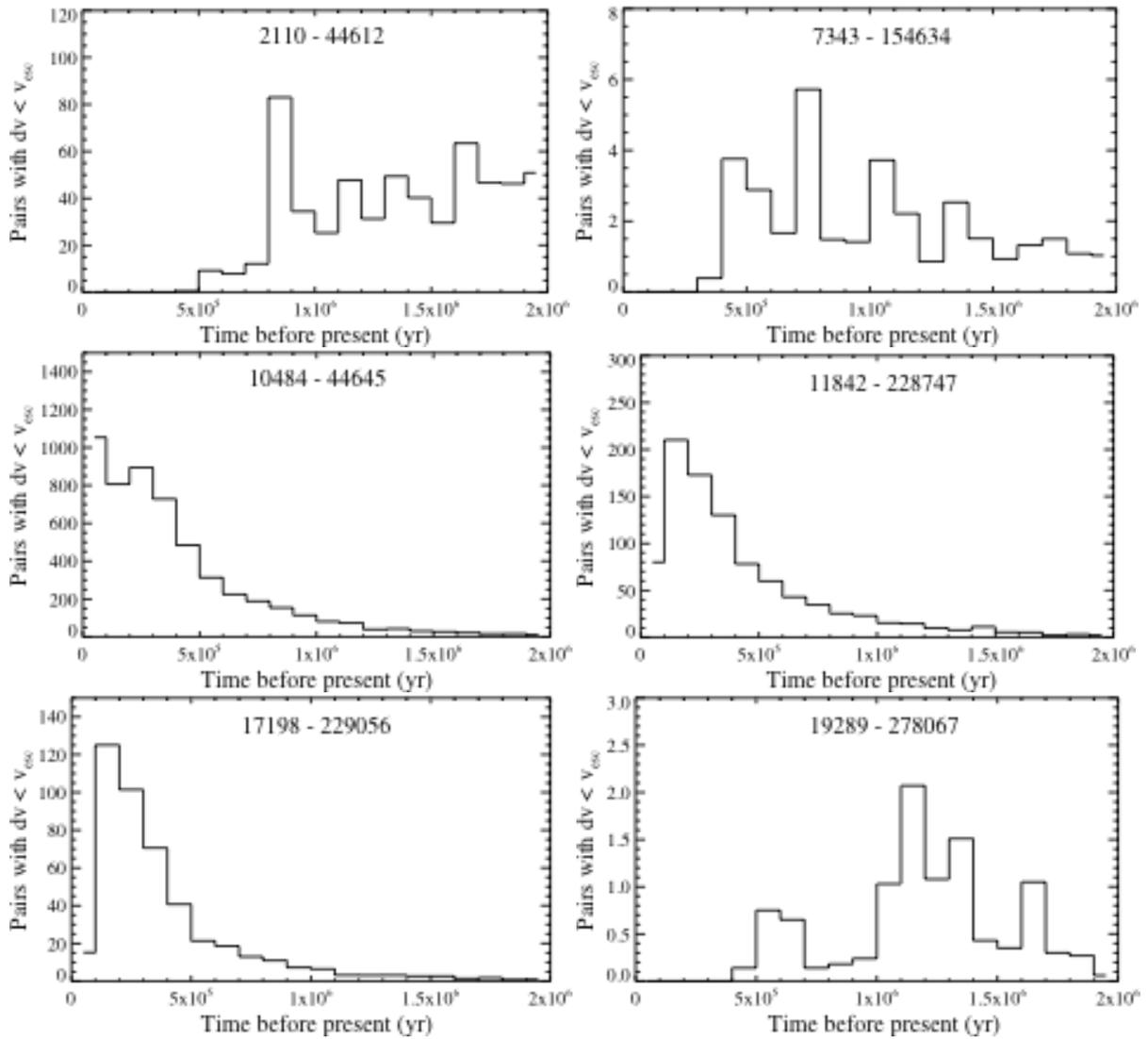

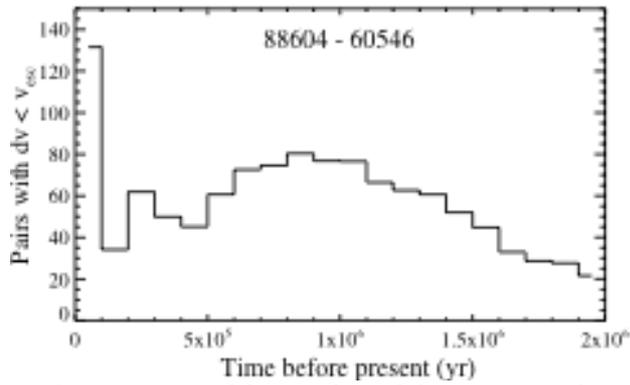

Fig. 8 Histograms of the number of clone pairings for which d$v$ < $v_{esc}$ during 2000 kyr integration

## 5. Discussion

Our results are summarized in Table 3. There is a large difference in our age estimate for 88604-60546 compared to Pravec et al (2010), who report a minimum age of 1000 kyr. Pravec et al. assigned this age as they found no convergences, defined by them as simultaneous d$v$ < $v_{esc}$ and d$r$ < $R_{hill}$, over 1000 kyr, whereupon their simulations stopped. As we noted earlier, after $dl$ is randomised, at 150 kyr in this case, it becomes increasingly difficult to track meaningfully the relative position of a specific clone pairing. Therefore the discrepancy in our age estimate can be explained due to our only requiring d$v$ < $v_{esc}$ convergence after $dl$ randomisation.

We can roughly say that a pair's spectrum is closer to its counterpart than a typical asteroid in that region if it ranks in the top half of possible pairings, i.e. 1-10. Five out of seven complete pairs meet this criterion, indicating that they are closer than typical. The spectral similarity of our observed pairs is evidence that they formed from a common parent body. In particular, 7343-154634 and 10484-44645 are remarkably close pairings. In contrast, 17198-229056 ranks 21/21 in its possible pairings with background asteroids, confirming the dissimilarity identified by Duddy et al. (2013). This is despite it having the lowest probability of being a coincidental pair in our sample, according to the technique of Pravec and Vokrouhlický (2009) ($P_2/N_p$ = 0.0005). Pair 19289-278067 is also revealed to have dissimilar spectra between its components (rank 17/21). The probability of 19289-278067 being a coincidental pair from population density considerations is higher ($P_2/N_p$ = 0.044), raising the question of whether the dissimilarity is attributable to not having a common parent body, i.e. being a spurious pair. In Table 3, we also estimate the mass ratio Δ$Q$, based on WISE diameters where available. 1979-13732 is the only pair here that breaches the 0.2 mass ratio barrier. However, its spectral similarity in comparison to background asteroids is very good (rank 2/21).

Moskovitz (2012) found one out of eleven asteroid pairs whose difference in colour formally fell outside the defined regime (Moskovitz applied a conservative generic uncertainty to a parameter extracted from *ugriz* SDSS photometry and to his own *BVRI* colours). From a comparison to the overall background SDSS MOC, performed by random drawing of nearest neighbours (these could be in different regions of the asteroid belt from the pairs), Moskovitz found that pair colours were more similar than background asteroid neighbours in 98% of trials. We find that two pairs have significantly different spectra compared to their neighbours, from a sample of seven pairs. We appear to find a greater proportion of dissimilar pairs, although a larger sample size is needed to explore this possible discrepancy.

However, there are subtle differences in the spectra that cannot be seen from a comparison of colours. The mean Q-type spectrum from DeMeo et al. (2009) has a longer wavelength Band I centre than the mean Sq-type. (88604) 2001 QH293 is a good match for the shape of the Sq-type spectrum, while (60546) 2000 EE85 is a better match for the Q-type, including a longer wavelength Band I centre. This might be due to a difference in composition, with (88604) more pyroxene-rich than (60546). Alternatively, it might indicate that primary surfaces are less weathered than secondaries, if we interpret that Sq-type are partially weathered Q-types (e.g. Binzel et al., 2004). In the spectral pairings where there is considerable difference, 17198-229056 and 19289-278067, both observed in visual wavelengths (Duddy et al. 2013), in each case the secondary has a much deeper one-micron absorption band, again suggesting a more weathered primary. We will consider the weathering state of pairs observed in or survey in more detail in further papers.

If they are real, small differences in spectral shape, are not evidence against the pairs being genetically related. There are other plausible physical causes such as: (1) patches across the surface meaning that the spectra change during rotation; (2) an original bi-lobed parent body with small compositional differences in the components; (3) one body having more weathered material on its surface as a consequence of the formation mechanism. This was discussed in more detail in the context of 17198-229056 in Duddy et al. (2013). We note that a group at MIT are also conducting a survey of asteroid pairs (Polishook et al. 2013). By comparing our data in cases where we have observed the same asteroid, we can hope to identify whether cause (1) is a significant factor in the near future. As we continue to gather more complete pairs in our own survey, we hope to identify trends in differences between the components' surfaces that will shed more light on formation mechanisms and let us distinguish between these possibilities.

**Table 3: Classification, physical properties, spectral similarity, and estimated ages for completed pairs**

| Vis/IR | Primary | Type | $H_V$ | $D_{eff}$ (km) | $p_V$ | Secondary | Type | $H_V$ | $D_{eff}$ (km) | $p_V$ | $\Delta Q$ | err. | $P_2/N_p$ | Rank /21 | Age (kyr) | err. |
|---|---|---|---|---|---|---|---|---|---|---|---|---|---|---|---|---|
| Vis | (1979) Sakharov | Sr | 13.4 | 4.22 ± 0.42 | 0.39 ± 0.08 | (13732) Woodall | Sr | 14.1 | 3.22 ± 0.34 | (0.39) | 0.44 | +0.38 -0.21 | 0.0118 | 2 | >2000 | |
| IR | (2110) Moore-Sitterly | S | 13.1 | 5.73 ± 0.57 | 0.16 ± 0.03 | (44612) 1999 RP27 | S/Sq/Q/K/L | 15.4 | 1.96 ± 0.39 | 0.20 ± 0.08 | 0.04 | +0.05 -0.02 | 0.0019 | 7 | >782 | |
| Vis | (7343) Ockegham | S | 13.8 | 4.11 ± 0.62 | 0.20 ± 0.06 | (154634) 2003 XX28 | S | 16.8 | 1.32 ± 0.21 | (0.20) | 0.03 | +0.05 -0.02 | 0.0593 | 3 | 982 | +880 -540 |
| IR | (10484) Hecht | V | 13.7 | 4.62 ± 0.55 | 0.23 ± 0.06 | (44645) 1999 RC118 | V | 14.6 | 2.07 ± 0.41 | 0.45 ± 0.18 | 0.09 | +0.14 -0.06 | 0.0057 | 1 | 348 | +825 -225 |
| Vis | (17198) Gorjup | Sa | 14.9 | | | (229056) 2004 FC126 | Sr | 17.5 | | | 0.03 | +0.09 -0.02 | 0.0005 | 21 | 285 | +775 -172 |
| Vis | (19289) 1996 HY12 | S | 15.3 | | | (278067) 2006 YY40 | Sq | 17.6 | | | 0.04 | +0.12 -0.03 | 0.0444 | 17 | 1189 | +529 -624 |
| IR | (88604) 2001 QH293 | S/Sq/Q/K/L | 13.1 | 5.80 ± 0.58 | 0.28 ± 0.06 | (60546) 2000 EE85 | S/Sq/Q/K/L | 14.4 | 3.31 ± 0.37 | (0.28) | 0.19 | +0.16 -0.09 | 0.0548 | 5 | 925 | +842 -754 |

Note: Visual spectra of pairs ("Vis") from Duddy et al. (2012, 2013). These have been reclassified, following the Bus taxonomy (Bus, 1999). We assign 10% uncertainty to $D_{eff}$ and 20% uncertainty to $p_V$ if the beaming parameter is measured by the WISE spacecraft, 20% and 40% respectively if not, unless the formal model uncertainty quoted in Masiero et al. (2011) is larger. $D_{eff}$ and $p_V$ for Ockegham from Duddy et al. (2012). $\Delta Q$ mass ratio is calculated assuming the same density of primary and secondary, and also the same albedo if only one object is detected by WISE (object albedo in parentheses). If neither pair has $D_{eff}/p_V$ measured, $\Delta Q$ is estimated from $H_V$, assuming ±0.5 mag. uncertainty. $H_V$ and $P_2/N_p$ values are from Pravec and Vokrouhlický (2009). Rank = Ranking of closeness of spectra compared to 20 other possible pairings in background population (see main text). Error on age or minimum limit is for 90% confidence interval (i.e. 5% of total distribution is below lower limit).

## Acknowledgements

Visiting Astronomer at the Infrared Telescope Facility, which is operated by the University of Hawaii under Cooperative Agreement no. NNX-08AE38A with the National Aeronautics and Space Administration, Science Mission Directorate, Planetary Astronomy Program. A part of this work was performed at the Jet Propulsion Laboratory under a contract with NASA. Taxonomic type results presented in this work were determined, in part, using a Bus-DeMeo Taxonomy Classification Web tool by Stephen M. Slivan, developed at MIT with the support of National Science Foundation Grant 0506716 and NASA Grant NAG5-12355. Part of the data utilized in this publication were obtained and made available by the MIT-UH-IRTF Joint Campaign for NEO Reconnaissance. The MIT component of this work is supported by NASA grant 09-NEOO009-0001, and by the National Science Foundation under Grants Nos. 0506716 and 0907766. Astronomical research at the Armagh Observatory is funded by the Northern Ireland Department of Culture, Arts and Leisure (DCAL). We thank an anonymous referee for many helpful comments that improved this manuscript.

## References

Binzel, R. P., Rivkin A. S., Stuart J. S., Harris A. W., Bus S. J., Burbine T. H. (2004). Observed spectral properties of near-Earth objects: results for population distribution, source regions, and space weathering processes. *Icarus*, 170(2), pp.259–294.

Bus S. J. (1999). Compositional Structure in the Asteroid Belt: Results of a Spectroscopic Survey, *PhD thesis*, MIT, Boston, MA.

Bus S. J. and Binzel R. P. (2002). Phase II of the Small Main-Belt Asteroid Spectroscopic Survey: The Observations, *Icarus* 158, 106-145.

Bus S. J. and Binzel R. P. (2002). Phase II of the Small Main-Belt Asteroid Spectroscopic Survey: A Feature-Based Taxonomy, *Icarus* 158, 146-177.

DeMeo F. E., Binzel R. P., Slivan S. M., Bus S. J. (2009). An extension of the Bus asteroid taxonomy into the near-infrared, *Icarus* 202, 160–180.

Duddy S. R., Lowry S. C., Wolters S. D., Christou A., Weissman P., Green S. F., Rozitis B. (2012). Physical and dynamical characterisation of the unbound asteroid pair 7343-154634. *A&A* 539, 36.

Duddy S. R., Lowry S. C., Christou A., Wolters S. D., Rozitis B., Green S. F., Weissman P. R. (2013). Spectroscopic observations of unbound asteroid pairs using the WHT, *MNRAS* 429, 63-74.

Masiero J. R., Mainzer A. K., Grav T. et al. (2011). Main Belt Asteroids With WISE/NEOWISE. I. Preliminary Albedos And Diameters, *ApJ* 741, 68-87.

Moskovitz N. A. (2012). Colors of dynamically associated asteroid pairs. *Icarus* 221, 63–71.

Polishook P., Moskovitz N., Binzel R. P., DeMeo F. (2013). 3rd Workshop on Binary Asteroids in the Solar System, Kohala Coast, Big Island of Hawaii, 30 June – 2 July 2013.

Pravec P. and Vokrouhlický D. (2009). Significance analysis of asteroid pairs, *Icarus* 204, 580–588.

Pravec P., Vokrouhlický D., Polishook D. et al. (2010). Formation of asteroid pairs by rotational fission, *Nature* 466, 1085–1088.

Rayner J. T., Toomey D. W., Onaka P. M., Denault A. J., Stahlberger W. E., Vacca W. D., Cushing M. C. (2003). SpeX: A Medium-Resolution 0.8–5.5 Micron Spectrograph and Imager for the NASA Infrared Telescope Facility. *PASP* 115, 362–382.

Reinsch C. H. (1967). Smoothing by Spline Functions, *Numerische Mathematik* 10, 177-183.

Rubincam D. (2000). Radiative Spin-up and Spin-down of Small Asteroids, *Icarus* 148, 2–11.

Scheeres D. J. (2009). Stability of the planar full 2-body problem, *Celest Mech Dyn Astr* 104, 103–128.

Vokrouhlický D. and Nesvorný D. (2008). Pairs of asteroids probably of a common origin, *AJ* 136, 280–290.

Vokrouhlický D. and Nesvorný D. (2009). The Common roots of asteroids (6070) Rheinland and (54827) 2001 NQ8, *AJ* 137, 111–117.

Walsh K. J., Richardson D. C., Michel P. (2008), Rotational breakup as the origin of small binary asteroids, *Nature* 454, 188–191.

Warner B. D., Harris A. W., Pravec P. (2009). The asteroid lightcurve database, *Icarus* 202, 134–146.

Willman M., Jedicke R., Moskovitz N., Nesvorný D. , Vokrouhlický D. , Mothé-Diniz T. (2010) Using the youngest asteroid clusters to constrain the space weathering and gardening rate on S-complex asteroids, *Icarus* 208, 758–772.

Xu S. (1994). CCD Photometry and Spectroscopy of Small Main-Belt Asteroids, *PhD thesis*, MIT, Boston, MA.

Xu S., Binzel R. P., Burbine T. H., Bus S. J. (1995). Small Main-Belt Asteroid Spectroscopic Survey: Initial Results, *Icarus* 115, 1-35.

Ye Q. Z. (2011). BVRI Photometry of 53 Unusual Asteroids, *AJ* 141, 32-39.